\documentclass[tradiabstract]{aa}
\usepackage{natbib}
\bibpunct{(}{)}{;}{a}{}{,} 

\usepackage{graphicx}
\usepackage{xcolor}
\usepackage{caption}
\usepackage{subcaption}
\usepackage{bm}
\colorlet{new}{red}
\usepackage{txfonts}
\begin{document}

\title{Fast formation of large ice pebbles after FU Orionis outbursts}
\titlerunning{Fast formation of large ice pebbles after FU Orionis outbursts}

\author{Katrin Ros\inst{1} \& Anders Johansen\inst{1,2}}
\authorrunning{Ros \& Johansen}

\institute{$^1$ Lund Observatory, Division of Astrophysics, Department of
Physics, Lund University, Box 118, 221 00 Lund, Sweden \\ $^2$ Center for Star and Planet Formation, Globe Institute,
University of Copenhagen, \O ster Voldgade 5-7, 1350 Copenhagen, Denmark\\e-mail:
\url{katrin.ros@astro.lu.se}}

\date{} 

\abstract{During their formation, nascent planetary systems are subject to FU Orionis outbursts that heat a substantial part of the disc. This causes water ice in the affected part of the disc to sublimate as the ice line moves outwards to several to tens of astronomical units. In this paper, we investigate how the subsequent cooling of the disc impacts the particle sizes. We calculate the resulting particle sizes in a disc model with cooling times between 100 and 1,000 years, corresponding to typical FU Orionis outbursts. As the disc cools and the ice line retreats inwards, water vapour forms icy mantles on existing silicate particles. This process is called heterogeneous nucleation. The nucleation rate per surface area of silicate substrate strongly depends on the degree of super-saturation of the water vapour in the gas. Fast cooling results in high super-saturation levels, high nucleation rates, and limited condensation growth because the main ice budget is spent in the nucleation. Slow cooling, on the other hand, leads to rare ice nucleation and efficient growth of ice-nucleated particles by subsequent condensation. We demonstrate that close to the quiescent ice line, pebbles with a size of about centimetres to decimetres form by this process. The largest of these are expected to undergo cracking collisions. However, their Stokes numbers still reach values that are high enough to potentially trigger planetesimal formation by the streaming instability if the background turbulence is weak. Stellar outbursts may thus promote planetesimal formation around the water ice line in protoplanetary discs.}



\maketitle
%

\section{Introduction}

Protoplanetary discs, the dusty and gaseous birthplaces of planets, are not stable and static environments. During the few million years in which the growth from dust to planets takes place, many physical parameters change. Very young stars experience strong outbursts that heat the disc, and these outbursts are followed by a phase of cooling. In this paper, we investigate how these ubiquitous heating events in young stellar systems influence the particle growth by shifting the water ice line outwards to up to tens of astronomical units.

In the quiescent stage of a typical protoplanetary disc, the water ice line is situated at 1-3 au. This corresponds to approximately 170 K, depending on the evolutionary stage of the disc \citep{martinlivio2012, bitschetal2015}. Beyond the ice line, water transitions from vapour to a solid form, allowing the solid density to double at this disc location \citep{hayashi1981, abodetal2019}. We thus expect the water ice line to potentially be an important location for growth towards planetesimals and planets. Therefore, understanding how particle growth takes place around the ice line is important in order to map the general growth from dust to larger bodies in the disc. Most works have so far been focused on a quiescent disc with a static ice line \citep[e.g.][]{stevensonlunine1988,cuzzizahnle2004, rosjohansen2013, schoonenbergormel2017, drazkowskaalibert2017}. However, the shifting of the ice line due to stellar outbursts will effectively reset the distribution of icy particles, and therefore, the consequences need to be investigated.

Early works that specifically modelled vapour condensation in the context of planet formation include \citet{stevensonlunine1988}, who investigated outwards vapour diffusion across the ice line and found significant growth of icy bodies there. However, this work ignored dynamical processes, such as inwards diffusion and radial transport of the resulting icy bodies, and they also assumed the instantaneous formation of new icy bodies without the need of a seed particle, followed by a rapid growth stage into planetesimals. \citet{cuzzizahnle2004} modelled the inwards drift of particles followed by sublimation and re-condensation of vapour on immobile sink particles. These works showed that the process of sublimation and condensation of vapour at the ice line has the potential to facilitate planetesimal growth there.

During the past decade, the dynamics of both vapour and particles were taken into account in modelling condensation and sublimation at the ice line by for example  \citet{rosjohansen2013, schoonenbergormel2017, idaguillot2016}.  These works showed dust growth to pebble sizes and pile-ups of particles due to sublimation and re-condensation of vapour at the ice line: As icy particles drift inwards past the ice line, they sublimate, and the vapour that is released can diffuse back across the ice line and re-condense onto particles there. It is nevertheless still debated whether the conditions for triggering planetesimal formation by the streaming instability, in terms of particle Stokes number and metallicity, can really be reached around the water ice line \citep{estradaumurhan2023}.

The outcome of condensation at the ice line is strongly affected by the number of particles onto which vapour can condense. A large particle number leads to vapour being distributed on a large surface area and thus a small net growth per particle, whereas fewer particles means that each individual particle experiences a larger growth. However, not the total number of solid particles around the ice line alone matters: The fraction of these particles that have an icy mantle is even more important. It is known from atmospheric chemistry that the condensation of vapour onto bare rocky particles, so-called  heterogeneous nucleation, is more difficult than condensation onto particles that are already covered by ice, so-called deposition \citep{vehkamaki2006}. The difference between heterogeneous nucleation and deposition was also confirmed in laboratory experiments \citep[see][and references therein]{hoosemohler2012}. This bottleneck for heterogeneous nucleation was not taken into account in some previous dust evolution models that included FU Orionis outbursts \citep{schoonenbergetal2017,hougekrijt2023}.
The presence of particles prone to heterogeneous nucleation and condensation growth was first suggested to be important by \citet{hubbard2017}. Assuming a set difference in the saturation ratio, that is, in the ratio of vapour pressure and saturated vapour pressure needed for condensation, between icy versus non-icy particles, the author found that this could lead to the growth of these icy particles to pebble sizes after the FU Orionis outburst at the expense of the non-icy particles, which stayed small.

In \citet{rosetal2019}, we modelled nucleation and deposition onto particles at a static water ice line using experimental temperature-dependent data for the saturation ratio required for condensation \citep{iracietal2010}. We found that the differentiation between icy and non-icy particles is indeed important in order to understand particle growth at the ice line. Growth takes place predominantly for already ice-covered particles because the vapour pressure outside the ice line is too low for heterogeneous nucleation (i.e. the formation of the first ice layer on an existing substrate) on bare silicate grains. This therefore leads to a dichotomy in which icy particles can grow to centimetre-sized pebbles in a narrow region around the ice line, while silicate particles stay dust-sized and diffuse out over the disc.

In this paper, we extend the work of \cite{rosetal2019} to the cooling phase after the disc has experienced a thermal heating event. The FU Orionis cooling timescale of 100--1,000 years is short enough for the dynamics of vapour and pebbles to be negligible {\citep{schoonenbergetal2017}}, but it is long enough for heterogeneous nucleation to become a rare event that only benefits a small subset of the silicate particles. In this way, we demonstrate the formation of centimetre-sized particles outside of the water ice line. These particles are large enough to trigger planetesimal formation by the streaming instability. An advantage to forming planetesimals in this way is that the pebble growth does not depend on the relatively poorly constrained diffusion rate of vapour across the ice line, in contrast to the prevalent model for planetesimal formation at the water ice line \citep{schoonenbergormel2017,drazkowskaalibert2017,rosetal2019,estradaumurhan2023}.

Our paper is organised as follows. In Section 2 we describe how we calculated the heterogeneous nucleation rate of ice on silicate dust, and we present our model for a protoplanetary disc that undergoes an FU Orionis outburst. We present the results of our calculation in Section 3. We discuss here that the condensation process is so fast that we can ignore physical effects such as collisions and sedimentation during the growth. In Section 4 we demonstrate that the ice particles grow large enough to potentially trigger planetesimal formation by the streaming instability. We briefly summarise our results and conclude in Section 5.

\section{Method}

\subsection{Nucleation calculations}

We discuss here the concept of heterogeneous nucleation, which means the formation of the first ice layer on a silicate particle, and how to calculate the nucleation rate. Nucleation onto silicate particles requires a significantly higher saturation than deposition (i.e. vapour-to-ice condensation) on grains that have already acquired an ice mantle. This is a key concept. A low nucleation rate therefore results in only a few but large icy particles, whereas a high nucleation rate allows many particles to grow, but only to comparatively smaller sizes. In order to understand the resulting size distribution, it is thus crucial to be able to model the nucleation rate correctly.

We calculated the nucleation rate using classical nucleation theory (CNT), which agrees well with experimental results when the typical prefactors are adopted \citep{hoosemohler2012, barahona2012}. 
In the following brief derivation, we differentiate between homogeneous nucleation, where water vapour nucleates into solid ice without a seed particle, and heterogeneous nucleation, where nucleation takes place on an existing solid seed particle. For water vapour at the ice line, only heterogeneous nucleation is relevant, but for simplicity, we here start from homogeneous nucleation in order to arrive at the expression for heterogeneous nucleation. For a more in-depth discussion of the various forms of nucleation, we refer to \cite{hoosemohler2012}. 

The energy barrier for the phase transition between vapour and ice is expressed in terms of the change in Gibbs free energy by formation of an ice particle of size $R$,
\begin{equation}
  \Delta G = \frac{4 \pi}{3} R^3 \Delta g_{\rm v} + 4 \pi R^2 \sigma\,,
\end{equation}
where $\sigma$ is the surface tension, and $\Delta g_{\rm v}$ is the change in free energy per volume for the formation of the solid phase. 

The critical cluster size beyond which an increase in $R$ leads to a reduction in free energy is found from setting $\delta \Delta G/\delta R=0$ to obtain
\begin{equation}
  R_{\rm c} = \frac{2 \sigma}{-\Delta g_{\rm v}}\,.
\end{equation}

The free energy difference can be expressed through the supersaturation factor $S=P_{\rm v}/P_{\rm sat}$, where $P_{\rm v}$ is the vapour pressure, and $P_{\rm sat}$ is the saturated vapour pressure,
\begin{equation}
  \Delta g_{\rm v} = (\rho_\bullet/\mu) k_{\rm B} T \ln S\,,
\end{equation}
 where $\rho_\bullet$ is the solid density, $\mu$ is the molecular mass of the water molecule, $k_{\rm B}$ is the Boltzmann constant, and $T$ is the temperature. The vapour pressure can be expressed using the ideal gas law,  
\begin{equation}
  P_{\rm v} = \frac{k_{\rm B} T}{m_{\rm v}}\rho_{\rm v},
\end{equation}
where $m_{\rm v}$ is the mass of a water molecule, and $\rho_{\rm v}$  is the density of vapour. The saturated vapour pressure is given by the Clausius-Clapeyron equation, and it yields
\begin{equation}
P_{\rm sat} = 6.034 \times 10^{11}\, {\rm Pa} \times {\rm e}^{-5938\,{\rm K}/T}
\end{equation}
using experimentally determined material constants for water \citep{haynesetal1992}. 

The energy barrier for homogeneous nucleation to reach the critical size is thus
\begin{equation}
  \Delta G_{\rm hom} = \frac{16 \pi \sigma^3}{3 (\Delta g_{\rm v})^2} = \frac{16
  \pi (\mu/\rho_\bullet)^2 \sigma^3}{3 (k_{\rm B} T \ln S)^2} \, .
\end{equation}

The rate at which clusters of the critical size form (expressed as per cubic metre per second) is
\begin{equation}
  J = J_0 {\rm e}^{-\Delta G_{\rm hom}/(k_{\rm B} T)}\,,
  \label{eq:Jhom}
\end{equation}
where $J_0$ is a kinetic pre-factor, $k_{\rm B}$ is the Boltzmann constant, and $T$ is the temperature \citep{kozasahasegawa1987}. This is the homogeneous nucleation rate, which gives the number of individual nucleation events per volume and the time for the case of the formation of new ice particles without a seed. In \cite{johansendorn2022}, we discussed the various forms that have been suggested for the pre-factor $J_0$.

For heterogeneous nucleation, we followed \cite{hoosemohler2012} and reformulated the change in Gibbs free energy as
\begin{equation}
  \Delta G_{\rm het} = \Delta G_{\rm hom} f \label{eq:DGhet},
\end{equation}
where the parameter $f(\theta)$ describes the wetting of the surface through
\begin{equation}
  f(\theta) = \frac{(2 + \cos \theta)(1-\cos \theta)^2}{4} \,.
  \label{eq:ftheta}
\end{equation}
Here, $\theta$ is the contact angle, which describes the flattening of the nucleating particle due to its coupling with the surface energy of the substrate. 

We can now generalise equation (\ref{eq:Jhom}) to the heterogeneous nucleation rate through
\begin{equation}
  j = j_0 {\rm e}^{-\Delta G_{\rm het}/(k_{\rm B} T)}\,.
  \label{eq:Jhet}
\end{equation}
Here, $j$ defines the number of nucleation events per particle surface area per second. For the heterogeneous kinetic pre-factor of equation (\ref{eq:Jhet}), we used
\begin{equation}
  j_0 = \frac{\alpha_{\rm c}}{\sqrt{f}} \frac{P_{\rm v}^2 (\mu/\rho)}{\mu k_{\rm
  B} T \nu_{\rm s}} \sqrt{\frac{\sigma}{k_{\rm B} T}} \exp \left( \frac{\Delta
  G_{\rm d}}{k_{\rm B} T} \right)
  \label{eq:prefac} \, .
\end{equation}
Here, $\alpha_{\rm c}$ is the sticking probability (assumed to be unity), $P_{\rm v} = n_{\rm v} k_{\rm
B} T$ is the vapour pressure, $\nu_{\rm s}$ is the lattice vibration frequency of
ice, and $\Delta G_{\rm d}$ is the desorption energy per molecule \citep{barahona2012}. The numerical values for the physical parameters needed to evaluate equation (\ref{eq:prefac}) are given in \cite{barahona2012}. Through equation (\ref{eq:Jhet}), we calculated the nucleation rate for a given temperature $T$ and super-saturation level $S$.

The contact angle $\theta$, introduced in Eq.\ \ref{eq:ftheta}, governs the ease with which nucleation takes place. For $\theta=\pi$ (no flattening), we recover homogeneous nucleation ($f=1$), where vapour does not benefit from nucleating on existing seed particles, but instead forms new ice particles. The other limit case where $\cos \theta=1$, giving $f=0$, corresponds to a scenario without any barrier to nucleation. Practically, this is equivalent to neglecting the bottleneck for heterogeneous nucleation, as was assumed in many earlier studies of vapour deposition in an astronomical context. Neglecting this bottleneck results in a small overall growth of each particle. We calculated the growth in radius, $\Delta R$, that each particle undergoes in a scenario where we compare the total particle mass including ice ($m_{\rm ice+rock}$) to the total rock mass alone ($m_{\rm rock}$) through
\begin{equation}
    \frac{m_{\rm ice+rock}}{m_{\rm rock}}=\frac{\sum_{\rm j= 1}^{\rm n} \frac{4\pi}{3} \left(R_{\rm j}^3 \rho_{\rm rock} + \left((R_{\rm j}+ \Delta R)^3 - R_{\rm j}^3 \right) \rho_{\rm ice}  \right)}
    {\sum_{\rm j=1}^{\rm n} \frac{4\pi}{3} R_{\rm j}^3 \rho_{\rm rock}} \, . 
\end{equation}
Here, $n$ is the number of particles, and $\rho_{\rm rock}$ and $\rho_{\rm ice}$ are the material densities of rock and ice, respectively. For $m_{\rm ice+rock}/m_{\rm rock}=2$, and using the particle size distribution described in Section \ref{s:growth}, we find that when heterogeneous nucleation is neglected, each particle grows equally in terms of radius by a very small amount, $\Delta R=5.8\,\mu {\rm m}$.

In our simulations with heterogeneous nucleation, we used a contact angle of $\cos \theta=0.83$, corresponding to $f=0.0085$, which is relevant for nucleation on silicate dust at a low temperature \citep{iracietal2010}. We performed additional experiments with $\cos \theta=0.9$, corresponding to $f= 0.0027$, and $\cos \theta=0.95$, corresponding to $f= 0.00066$. The first case simply reduced the region of large pebbles outside of the ice line, while the second case substantially suppressed the overall particle growth as the rate of heterogeneous nucleation increased dramatically. This agrees with the findings from the limit case where $f=0$.

\subsection{Disc model and cooling}

We implemented a simple 1D model of a very young protoplanetary disc that accretes gas onto the star at a rate of $\dot{M} = 10^{-7}\,M_\odot\,{\rm yr^{-1}}$. We assumed that the accretion rate is steady throughout the inner regions of the protoplanetary disc. This allowed us to calculate the gas column density $\varSigma$ from the flux expression $\dot{M} = 3 \pi \nu \varSigma$. Here, $\nu = \alpha c_{\rm s} H$ is the viscosity of the disc, $c_{\rm s}$ denotes the sound speed and $H = c_{\rm s}/\varOmega$ the gas scale height ($\varOmega$ is the Keplerian frequency at distance $r$). We followed \cite{idaetal2016} and set the disc temperature to
\begin{equation}
  T = 150\,{\rm K}\left(\frac{L_\star}{L_\odot}\right)^{2/7} \left( \frac{r}{\rm au}\right)^{-3/7} \, .
\end{equation}
Here, $L_\star$ is the luminosity of the star. We ignored viscous heating here because this is likely inefficient if the angular momentum transport is dominated by disc winds \citep{morietal2021}.

During FU Orionis outbursts, young stars reach luminosities of up to $10^2$ -- $10^3$ $L_\odot$ \citep{hartmannkenyon1996}. These outbursts likely originate in the innermost regions of the protoplanetary disc \citep{zhuetal2007,zhuetal2009, hillenbrandfindeisen2015,pignataleetal2018} and decay on timescales of decades to centuries. We focused on large outbursts, with $L=10^3\,L_\odot$, and on two values of the cooling timescale ($100$ and $1,000$ years). The protoplanetary disc model was initialised in its high-luminosity state and was subsequently exponentially cooled down to its pre-outburst equilibrium temperature.

We define the nucleation timescale for a particle as
\begin{equation}
    t_{\rm nuc} = \frac{1}{j\, A_{\rm sur}}\,,
\end{equation}
where $A_{\rm sur}$ is the surface area of the seed particle, and $j$ is given by equation (\ref{eq:Jhet}). This shows that the nucleation timescale scales with particle size as $t_{\rm nuc} \propto R^{-2}$.

\begin{figure}
  \begin{center}
    \includegraphics[width=\linewidth]{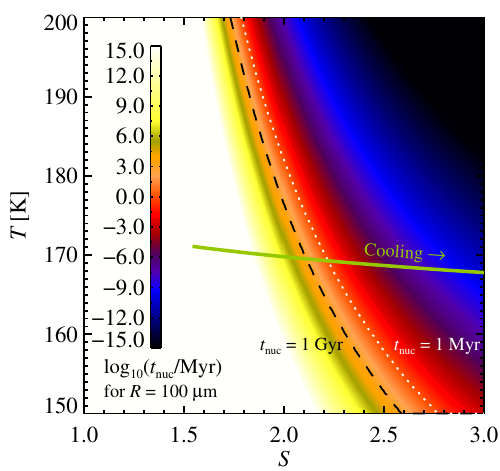}
  \end{center}
  \caption{Nucleation timescale for a single 100-micrometre-sized grain for a range of super-saturations ($S$) and temperatures ($T$). The dotted line marks the 1 Myr nucleation timescale, and the dashed line shows the 1 Gyr nucleation timescale. We illustrate a schematic cooling path at the conditions of the water ice line with the green line.}
  \label{fig:nucl_rate}
\end{figure}

To illustrate the rate of heterogeneous nucleation, we show in Fig.\ \ref{fig:nucl_rate} the nucleation timescale for a particle of size $R=100\, \mu \rm m$ over a range of super-saturations and temperatures, calculated for the conditions at the water ice line.  The figure clearly shows that the nucleation rate is a very steep factor of supersaturation. For all temperatures shown here, the nucleation timescale is extremely long for $S\leq2$. This means that nucleation effectively does not occur until $S\geq2$, at which point the cooling path of a particle crosses a steep nucleation cliff.

The cooling path of a particle at typical conditions at the water ice line is schematically illustrated with a green line. The cooling path is the same regardless of the cooling timescale, but the outcome in terms of the resulting number of ice-nucleated particles and their size changes depends on how fast the disc cools. When a disc cools quickly, the nucleation cliff is also climbed quickly, meaning that many particles are able to nucleate. The resulting particle sizes are therefore smaller. In the opposite case, where the cooling is slow, fewer particles nucleate and grow to larger sizes because the particles climb the nucleation cliff more slowly. 

\subsection{Particle growth}
\label{s:growth}

In \cite{johansendorn2022}, we calculated the homogeneous nucleation of refractory minerals in the innermost regions of the protoplanetary disc after an FU Orionis outburst. Here, we focus instead on heterogeneous nucleation of water ice farther away from the star. Heterogeneous nucleation has the added complexity that only a very small fraction of the silicate particles nucleate ice. We therefore followed two populations of particles: (i) the silicate particles that may experience heterogeneous nucleation, and (ii) the rare ice-nucleated particles that grow directly by vapour deposition.

We assumed that the disc is populated by silicate particles with a minimum size of 1 $\mu$m and a maximum size of 1 mm. We ignored smaller dust sizes because their high curvature will likely prevent the formation of a stable ice layer \citep{kuroiwasirono2011}. The maximum size is motivated by models of dust coagulation \citep{guttleretal2010,zsometal2010} and by observational constraints on pebble sizes in protoplanetary discs \citep{carrasco-gonzalezetal2019}. Any solid ice in the modelled part of the disc was assumed to have sublimated during the outburst and to have left behind water vapour and either intact or disrupted silicate aggregates \citep{spadacciaetal2022,aumatellwurm2011}. The dust number $N$ as a function of size is distributed as a power law ${\rm d}N/{\rm d}R \propto R^{-q}$ with index $q=3.5$. These particles, as well as the water vapour, were assumed to be mixed with the gas and not sedimented to the mid-plane. We discretised the silicate size range into 12  logarithmically spaced super-particles populating each of the 400 radial annuli that span from 0.4 au to 30 au. During the cooling of the disc, we kept track of the temperature $T$ and super-saturation level $S$ in each annulus. In every time step, we then calculated the number of physical particles in each super-particle, $\Delta N_i^{({\rm ice})}$, that underwent heterogeneous nucleation.

Eq.\, (\ref{eq:Jhet}) gives us the rate of nucleation per unit area. This yields $\Delta N_i^{\rm (ice)} = j \, 4 \pi R_i^2 \, N_i \, \Delta t$ new ice particles, with $R_i$ denoting the radius of physical particles in a super-particle of index $i$, $\Delta N_i$ denoting the total number of physical particles in the super-particle in each time step, and $\Delta t$ denoting the time step. These ice-covered particles were transferred to a new super-particle whose physical particles, initialised at the critical cluster size, were allowed to undergo direct depositional growth. In this way, the number of super-particles increased until the vapour reservoir was depleted. 

After a particle has nucleated, it will grow purely by deposition of vapour. The growth rate by deposition for a spherical particle is 
\begin{equation}
  \dot{R} = \frac{v_{\bot}}{\rho_\bullet} (\rho_{\rm v}- \rho_{s})\,.
\label{eq:deprate}
\end{equation}
Here, $\rho_\bullet$ is the material density of water ice, $\rho_{\rm v}$ is the water vapour mass density, and $\rho_{\rm s}$ is the saturated vapour density of water at the environmental temperature. We did not include dust smaller than 1 micron, and we therefore ignored that $\rho_{\rm s}$ increases for very small dust sizes \citep{kuroiwasirono2011}.
The water vapour flux speed is 
\begin{equation}
  v_{\bot} = \sqrt{ \frac{k_{\rm B} T}{2 \pi m_{\rm v} }}, 
\end{equation}
where $k_{\rm B}$ is the Boltzmann constant, $T$ is the temperature, and $m_{\rm v}$ is the water molecule mass. 


\section{Results}

\subsection{Pebble growth after outburst}

\begin{figure}
  \begin{center}
    \includegraphics[width=\linewidth]{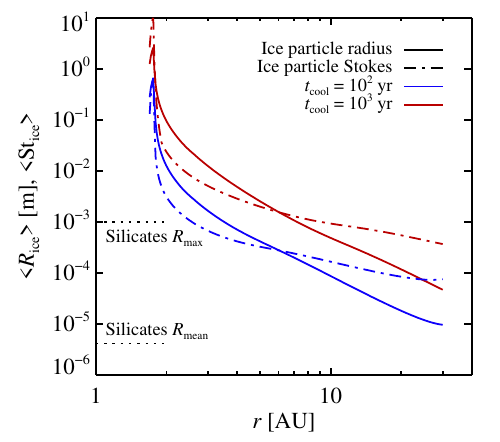}
  \end{center}
  \caption{Mean sizes and Stokes number for particles after an FU Orionis outburst with a decay time of 100 yr in blue and 1,000 yr in red. The sizes of silicate particles at the start of the simulation are shown as dotted lines, with a maximum size of one millimetre. Growth to the largest particles occurs just outside of the quiescent ice line.}
  \label{fig:het_nuc}
  \end{figure}

In Fig.\ \ref{fig:het_nuc} we compare the mean particle sizes after an FU Orionis outburst for two different decay times, 100 and 1,000 years. For both decay times, growth to the largest sizes is achieved just outside of the quiescent ice line, where the cooling stops at a super-saturation just above unity. However, the cooling timescale does have a significant impact on the resulting particle sizes. For a fast cooling timescale, $t_{\rm cool}=100\, {\rm yr}$, a large number of grains nucleates. The vapour is therefore spread out over a large number of particles, resulting in a smaller growth for each individual particle. For a slow cooling timescale, $t_{\rm cool}=1,000\, {\rm yr}$, we observe fewer nucleation events. The icy particles that do form by nucleation can grow to much larger sizes in this case. 

\begin{figure}
  \begin{center}
    \includegraphics[width=\linewidth]{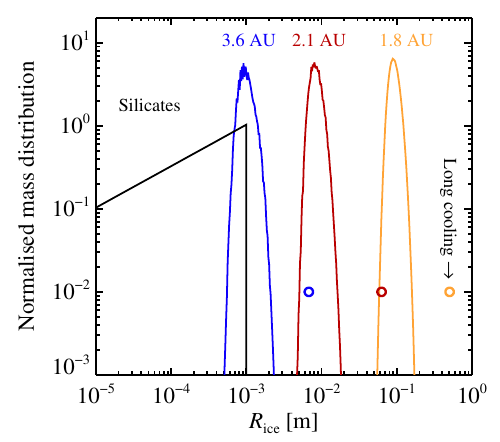}
  \end{center}
  \caption{Mass distribution of particles (normalised such that the integral over the logarithmic radius is always unity) just outside of the ice line at 1.8 au (yellow line), 2.1 au (red line), and 3.6 au (blue line) after the ice line has returned to its quiescent state in a scenario with a cooling time of $t_{\rm cool}=100\, {\rm yr}$. We indicate the particle size for the longer cooling time of $1,000\, {\rm yr} $ with circles. The black line shows the mass distribution of bare silicate particles.}
  \label{fig:size_dist}
  \end{figure}

The size distribution of icy particles is shown in Fig.\ \ref{fig:size_dist} for representative locations in the disc. At each location (1.8 au, 2.1 au, and 3.6 au), the mass is concentrated in a narrow size range with particle sizes of about 10 cm just outside the quiescent ice line at 1.8 au, 1 cm at 2.1 au, and 1 mm at 3.6 au for the short cooling time of $t_{\rm cool}=100\, {\rm yr}$. For a longer cooling time of $t_{\rm cool}=1,000\, {\rm yr}$, the particle sizes are even larger. The figure also shows the initial size distribution of silicate particles, as a comparison.

The evolution of a representative particle nucleating close to the ice line is shown in Fig.\ \ref{fig:pebble_evo}. Particles predominantly nucleate during the short time when supersaturation (shown in panel a) is about $S=2$. Nucleation can be seen as an instantaneous event. As shown in panels b) and c), the continued growth to pebbles with ${\rm St}\approx0.01$ by vapour deposition also occurs quickly, on a timescale of about ten years. We can therefore assume that nucleation and depositional growth in this early growth phase take place without being influenced by other processes such as sedimentation and collisions. We examine this assumption more closely in the following sections.
  
\subsection{Particle collisions}

In our model, we ignored collisions that could take place during the period of condensation growth. The condensed water molecules form layers of solid ice on the ice-nucleated particles. Macroscopic particles of solid ice, as opposed to aggregates made of microscopic monomers, are not expected to stick to each other at the relative speeds experienced in the protoplanetary disc \citep[see][and references therein]{johansenetal2014}. However, solid ice particles may experience cracking and destruction in relatively low-speed collisions.

\begin{figure}
 \begin{center}
  \includegraphics[width=\linewidth]{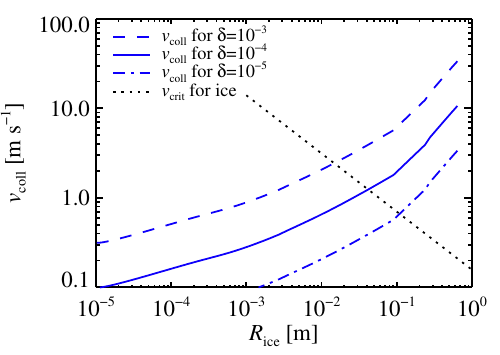}
  \end{center}
  \caption{Ice particle collision speed as a function of the mean ice particle size in each radial bin for a turbulent diffusion coefficient of $\delta=10^{-3}$, $\delta=10^{-4}$, and $\delta=10^{-5}$. The critical velocity for ice fragmentation is over-plotted as a dotted line. Ice particles can grow to a few centimetres before they risk fragmentation in collisions.}
  \label{fig:coll_speed}
\end{figure}

In Fig. \ref{fig:coll_speed} we plot the collision velocity as a function of particle size for three different turbulence levels, $\delta=10^{-3}$, $\delta=10^{-4}$, and $\delta=10^{-5}$ ($\delta$ denotes the dimensionless turbulent diffusion coefficient, defined equivalently to the $\alpha$-viscosity coefficient as the turbulent diffusion coefficient normalised by $c_{\rm s} H$). We calculated the collision speed from the expression valid for two equal-sized particles, 
\begin{equation}
  v_{\rm c} = \sqrt{3} \sqrt{\rm St} \sqrt{\delta} c_{\rm s}
  \label{eq:vc}
\end{equation}
{\citep{ormelcuzzi2007}}. Here, $c_{\rm s}$ is the sounds speed of the gas, and ${\rm St}$ is the dimensionless Stokes number, which quantifies the strength of coupling between gas and dust and is defined as
\begin{equation}
   {\rm St}= \frac{\pi}{2} \frac{R \rho_\bullet}{\varSigma} \label{eq:St}
\end{equation}
in the Epstein drag regime \citep{weidenschilling1977}. Here, $\varSigma$ denotes the gas surface density. From Eq. (\ref{eq:St}), we can describe how the Stokes number scales with particle size,
\begin{equation}
   {\rm St} \sim 0.0015 \left( \frac{R}{\rm cm} \right) \left(\frac{r}{2\,{\rm au}} \right)^{-15/14} \, ,
\end{equation}
where $r$ is the distance from the central star. We assumed a gas accretion rate $\dot{M}_\star= 3 \pi \varSigma \alpha c_{\rm s} H = 10^{-7}\,{\rm M_\odot}\,{\rm yr}^{-1}$ when calculating the gas surface density $\varSigma$.

We also show in Fig. \ref{fig:coll_speed} the critical fragmentation velocity, where icy particles start to fragment in collisions \citep{hauketal2015}. A comparison of the collision speeds to the critical fragmentation velocity shows that  icy particles reach at least centimetre-sizes without being hindered by fragmentation for all turbulence levels.
Any growth beyond these sizes could nevertheless be impacted by fragmenting collisions. To understand the potential impact of collisions, we now compare the condensation growth timescale to the collision timescale. 

The collision timescale can be written as $\tau_{\rm col} = 1/(n \sigma v_{\rm col})$, where $n$ is the number density, and $\sigma$ is the collision cross section. Inserting the collision speed from equation (\ref{eq:vc}), we arrive at the expression
\begin{equation}
\tau_{\rm col} = \left(\frac{300}{\varOmega}\right) \left(\frac{\rm St}{0.01}\right)^{0.5} \left(\frac{\rho_{\rm p}/ \rho_{\rm g}}{0.01}\right)^{-1} \left(\frac{\delta}{10^{-4}}\right)^{-1/2}, 
\end{equation}
where $\rho_{\rm p}$ and $\rho_{\rm g}$ are the mid-plane dust and gas density. For centimetre-sized particles with ${\rm St} = 0.001$, we then obtain a collision timescale of $\tau_{\rm col}\approx 40 \,\rm {yr}$ and a timescale of $\tau_{\rm col}\approx 135 \,\rm {yr}$ for decimetre-sized particles with ${\rm St} = 0.01$.

We compared this to the growth timescale of a particle, defined as
\begin{equation}
  \tau_{\rm gro} = \frac{R}{\dot{R}} \, ,
\end{equation}
where the growth rate by deposition for spherical particles is given by Eq.\ \ref{eq:deprate}. 

\begin{figure}
 \begin{center}
\includegraphics[width=\linewidth]{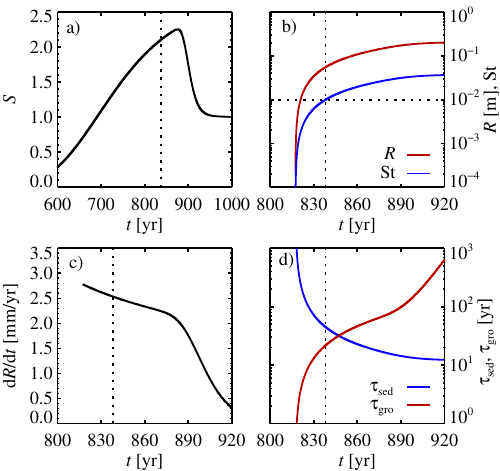}
  \end{center}
  \caption{Evolution of a representative pebble at 1.8 au, nucleating when $S=2$, corresponding to a time of $ t \approx 840\,{\rm yr}$. Panel a shows the supersaturation $S$ at 1.8 au. Panel b shows the particle size in metres and Stokes number. We mark ${\rm St}=0.01$ with a dotted line. Panel c shows the growth rate in millimetres per year, and panel d shows the sedimentation and growth timescales. The cooling time of the disc is $t_{\rm cool}=100\,{\rm yr}$.}
  \label{fig:pebble_evo}
\end{figure}

The saturated vapour pressure is $\rho_{\rm s} \ll \rho_{\rm v}$ during nucleation because ice nucleation occurs when the supersaturation is $S > 2$. We therefore write the condensation rate as
\begin{equation}
  \dot{R} = 3.5\,{\rm mm\,yr^{-1}}\,\left( \frac{\rho_{\rm v}}{10^{-9}\,{\rm kg\,m^{-3}}} \right) \left( \frac{T}{170\,{\rm K}} \right)^{1/2} \, .
\end{equation}
Here, $\rho_{\rm v} = 10^{-9}\,{\rm kg\,m^{-3}}$ represents the characteristic vapour density at 2 au from the distance from the star and an accretion rate of $\dot{M}=10^{-7}\,M_\odot\,{\rm yr}^{-1}$. This estimate corresponds well to the actual deposition rate reported in panel c of Figure \ref{fig:pebble_evo}. The growth timescale follows from $R/\dot{R}$ as
\begin{equation}
    \tau_{\rm gro} \approx 0.3 \,{\rm yr} \, \left( \frac{R}{1\,{\rm mm}} \right) \left( \frac{\rho_{\rm v}}{10^{-9}\,{\rm kg\,m^{-3}}} \right)^{-1} \left( \frac{T}{170\,{\rm K}} \right)^{-1/2}
      \label{eq:tau}\, .
\end{equation}
The collision timescale is thus several times longer than the growth timescale for millimetre-sized and centimetre-sized particles. This comparison thus shows that the frequency with which these collisions can occur during the growth phase is low, and it would therefore not impact our results significantly. However, collisions will be important during the subsequent sedimentation phase.

\subsection{Sedimentation}

We compared the growth timescale to the sedimentation timescale to show that our assumption that the particles are located in the full gas scale-height layer rather than in a dense sedimented mid-plane layer is valid for the duration of their growth process.

The timescale to sediment to the mid-plane of the protoplanetary disc is
\begin{equation}
    \tau_{\rm sed} \sim \frac{1}{\varOmega \, {\rm St} } \, ,
\end{equation} \citep{youdinlithwick2007} where $\varOmega$ is the orbital frequency. This gives
\begin{equation}
    \tau_{\rm sed} = 450 \, {\rm yr} \, \left( \frac{\rm St}{0.001} \right)^{-1} \left( \frac{r}{2\,{\rm au}} \right)^{3/2} \,.
\end{equation}
Comparing this to the growth timescale in Eq.\ (\ref{eq:tau}), we find that with a difference of two orders of magnitude at 1 cm (${\rm St}=0.001$) and a similar magnitude at 10 cm (${\rm St}=0.01$),  we can conclude that the sedimentation timescale is too long for sedimentation to be important for growth by deposition. We therefore did not include sedimentation of particles in our simulations. However, settling will become important shortly after the deposition growth phase.

\section{Implications for planetesimal formation}  

An important question for understanding the role of heterogeneous ice nucleation for planet formation is whether and how icy pebbles that formed by nucleation and deposition of vapour can grow further into planetesimals after they have sedimented to the mid-plane. Here, we place our results in the context of the leading mechanism of planetesimal formation, the streaming instability, in which particles clump together due to gas drag felt by the solids to form clumps that are dense enough for them to gravitationally collapse into solid planetesimals \citep{youdingoodman2005,johansenetal2007}.

\begin{figure}
  \begin{center}
    \includegraphics[width=\linewidth]{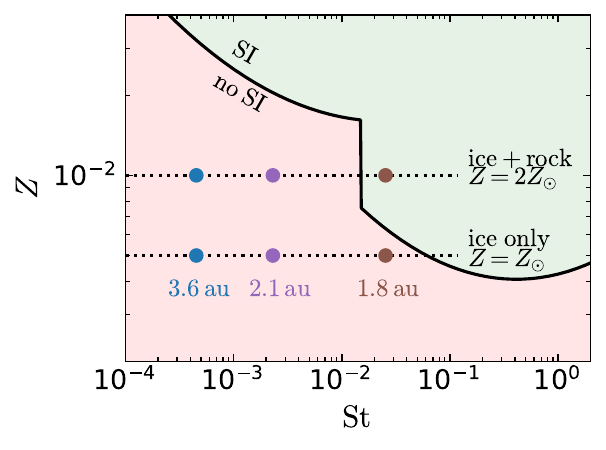}
  \end{center}
  \caption{Threshold for particle clumping via the streaming instability for different particle stopping times and metallicities. The green area in the upper right part denotes the parameter space within which clumping via streaming instability is possible. The filled coloured circles show our data for different disc locations with a cooling time of $t_{\rm cool}=100\, {\rm yr}$. The lower row corresponds to a disc with solar metallicity within which the interacting solids are made of ice, and the upper row is either a high-metallicity disc or a disc with solar metallicity, where both ice and rock contribute to clumping.}
  \label{fig:si}
\end{figure}

In Fig.~\ref{fig:si} we plot the threshold for clumping via the streaming instability for a range of particle Stokes numbers and metallicities using the fit to computer simulations from \citet{liyoudin2021}. A combination of a high enough metallicity and particle Stokes numbers gives the parameter space within which clumping via streaming instability is possible. This area is the upper region in the figure, and the lower region corresponds to the parameter space within which clumping via streaming instability is not possible. 

For a disc with with solar metallicity ($Z=0.005$), where we only allowed water ice to contribute to the clumping, the particles are not large enough to trigger the streaming instability in our simulations. However, for the scenario in which the disc, either globally or locally, has a higher metallicity than $Z=0.01$ or where both rock and ice are included as solids that can contribute to clumping (e.g. if the silicate dust aggregates collide with and adhere to the ice particles) the outcome is more promising. In this case, the largest particles that can reach centimetre-sizes just outside of the ice line fall in the range in which clumping via streaming instability is possible. This would also change the chemical composition of the planetesimals from pure ice (if formed only out of the ice particles that grew by condensation) to a mixture of silicates and ice. 

We emphasise that we did not model the step from particle growth through sedimentation and to clumping by the streaming instability, but rather compared the particle sizes resulting from our model to typical theoretical limits for planetesimal formation through the streaming instability \citep{liyoudin2021,limetal2023}. 

Clumping by streaming instability is expected to take place in the mid-plane. The ratio of the solids to gas density here can be calculated for a specific particle size from the overall metallicity, $Z$, and the dimensionless turbulent diffusion
coefficient, $\delta$, as 
\begin{equation}
\begin{aligned}
 \left(\rho_{\rm p}/\rho_{\rm g} \right)_{\rm mid} &=
 Z \sqrt{ \frac{{\rm St} + \delta}{\delta} } \sim Z \sqrt{ \frac{\rm St}{\delta}} \\
 &=  0.3 \, \left(\frac{Z}{0.01}\right) \, \left( \frac{\rm St}{0.01}\right)^{0.5} \, \left( \frac{\delta}{10^{-5}}\right)^{-0.5} 
 \end{aligned}
\end{equation}
\citep{johansenetal2014}.
For a disc with a low turbulence of $\delta=10^{-5}$ and solar metallicity in which both rock and ice contribute to clumping, we find a mid-plane ratio of solids to gas of 0.3, corresponding to the value in the simulations by \citet{liyoudin2021}. Newer simulations by \citet{limetal2023} explored the threshold for streaming instabilities for a range of values of $\delta$ only down to $10^{-4}$, but the trend of their simulations indicates that $\delta=10^{-5}$ would lead to clumping for values corresponding to our resulting particle sizes. A low turbulence like this would also decrease the effect of collisional fragmentation, as is clear from Figure \ref{fig:coll_speed}.

We note that the turbulence level of $\delta=10^{-5}$ that we adopted here is lower than in many simulations, and that a high background diffusion would restrict the effectiveness of clumping by the streaming instability \citep{umurhanetal2020}. However, there is observational support for a low level of turbulence \citep{villenaveetal2022}. Theoretically, we also expect the mid-plane layer to be less turbulent than the upper disc layers, as it has been shown that the vertical shear instability stirs the upper layers of the disc, but not the mid-plane \citep{schaferetal2020, schaferjohansen2022}.

Finally, we need to assess the amount of material available for clumping into planetesimals at the ice line. The mass in the annulus where particles grow above the streaming instability threshold can be estimated in Earth masses, $M_{\rm E}$, as 
\begin{equation}
\begin{aligned}
    M &= 2 \pi \, r \, W \,\Sigma_{\rm peb} \\&\approx 0.4 M_{\rm E} \, \left(\frac{r}{1.8\,{\rm au}}\right) \, \left( \frac{W}{0.1\,{\rm au}}\right) \, \left(\frac{\varSigma_{\rm g}}{\rm 10^4 \,kg/m^2}\right) \left( \frac{Z}{0.01} \right)\,,
    \end{aligned}
\end{equation}
where $\Sigma_{\rm peb}$ is the surface density of pebbles, and $W$ is the width of the annulus. Estimating the width of the annulus to $W=0.2 \,{\rm au}$, and assuming that both ice and rock contributes to the clumping, we thus determine a mass of the material available for clumping of $M=0.8\,M_{\rm E}$. This confirms the potential for a pathway towards planets around the ice line.

\section{Conclusions}

We have investigated the effects of stellar outbursts, specifically, FU Orionis outbursts, and the subsequent cooling of the disc on heterogeneous ice nucleation onto silicate particles and particle growth by deposition. During an outburst, the temperature in the disc increases dramatically, causing the ice line to shift outwards and to sublimate all the ice in the affected region. The subsequent cooling phase allows the ice line to move back to its quiescent position, and vapour is then redistributed onto solid grains. Since nucleation of ice from vapour onto bare grains requires a higher supersaturation than the deposition of vapour onto existing ice grains, this results in the condensation of ice on a few particles that can then grow to be very large. Our main findings are summarised below.
\begin{enumerate}
  \item {Particle growth:} A stellar outburst resets the particle size distribution so that particle growth starts over from bare silicate grains. Beginning from a particle distribution with a maximum size of $R=1\, {\rm mm}$, we find resulting sizes of several centimetres just outside the quiescent ice line. Growth nevertheless affects a broad region between $r_{\rm in}\approx 2 \, \rm au$ and $r_{\rm out}\approx 10 \, \rm au$. Importantly, growth by condensation following heterogeneous nucleation is several times faster than growth by coagulation, meaning that it can be less dependent on typical collisional growth barriers. \\
  \item {Cooling time dependence:} The timescale on which the disc cools is important for the resulting particle sizes. A slower cooling leads to larger particles. However, for both the long and short cooling timescales that we investigated, we find growth to at least centimetre-sized pebbles outside of the ice line. The width of the growth zone and the maximum particle size are nevertheless both larger in the case of slower cooling. For the longer cooling time investigated in this paper, $t_{\rm cool}=1,000\, \rm{yr}$, centimetre-sized particles form out to 3 au, whereas the shorter cooling time of $t_{\rm cool}=100\, \rm{yr}$ results in growth to at least centimetre-sized particles out to 2 au. \\
  \item {Collisions and sedimentation:} We find that both collisions and sedimentation can be ignored in the modelling of nucleation and deposition in the aftermath of an FU Orionis outburst because of the rapid condensational growth that follows heterogeneous nucleation. Collisions between particles of a few centimetres and larger could nevertheless lead to fragmentation for typical turbulence levels after the phase of condensational growth. \\
  \item {Relevance for the streaming instability:} After an outburst, we find particles of several centimetres outside the ice line. This corresponds to Stokes numbers as high as ${\rm St} = 0.01-0.02$, which is sufficient to trigger the streaming instability for a metallicity of $Z=10^{-2}$ in a low-turbulence disc ($ \bm{\delta}=10^{-5}$). We did not model the actual step from particle growth through sedimentation and to the streaming instability, but we tentatively find that stellar outbursts lead to favourable conditions for growth into planetesimals. \\
  \item {Importance of nucleation:} Heterogeneous nucleation is a crucial component in particle growth by ice condensation. The final particle sizes depend on how many particles are available for vapour deposition (typically referred to as condensation in astronomical literature), with larger particles being favoured in a scenario in which nucleation is rare. We have demonstrated that the aftermath of a stellar outburst with relatively slow cooling ($100\, {\rm yr}\lesssim t_{\rm cool}\lesssim 1000\, \rm{yr}$) provides just such favourable conditions as needed for particle growth. Importantly, the condensation growth that follows nucleation is so rapid that dynamical processes such as diffusion, sedimentation, and drift have little or no influence on the ensuing particle sizes.
\end{enumerate}
We did not model the collisional evolution of the size distribution of silicate particles and solid-ice particles after the outburst. The largest ice particles will collide fast enough to experience cracking collisions, which will reduce their size to a few centimetres at most. At the same time, silicate dust aggregates colliding with solid-ice particles may adhere their monomers to the icy surface. Laboratory experiments on the outcome of collisions between dust aggregates and macroscopic solid water ice particles are lacking in the literature so far. Additional modelling that takes the coexistence of dust aggregates and solid ice into account as well as laboratory experiments are therefore needed to better understand particle growth and planetesimal formation outside the water ice line after FU Orionis outbursts.


\begin{acknowledgements}
We thank the referee for a thorough reading of the manuscript and for providing comments and questions that helped improve the quality of this paper. A.J.~acknowledges funding from the European Research Foundation (ERC Consolidator Grant 724687-PLANETESYS), the Knut and Alice Wallenberg Fountdation (Wallenberg Scholar Grant 2019.0442), the Swedish Research Council (Project Grant 2018-04867), the Danish National Research Foundation (DNRF Chair Grant DNRF159) and the Göran Gustafsson Foundation.
\end{acknowledgements}

%
%

\bibliographystyle{aa} 
\bibliography{refs} 


\end{document}